\documentclass[12pt]{article}

\usepackage{a4wide}

\usepackage{amsmath}
\usepackage{cite}
\usepackage{epsfig}
\usepackage{color}
\usepackage{graphics}
\usepackage{amssymb}

\newcommand{\cnf}{\emph{cf.}\ }
\newcommand{\ie}{\emph{i.e.}\ }
\newcommand{\eg}{\emph{e.g.}\ }
\newcommand{\as}{\alpha_{\mathrm{s}}}
\newcommand{\ee}{e^+e^-}
\newcommand{\cF}{{\cal{F}}}    %   the multiple emission correction factor
\newcommand{\muf}{\mu_\textsc{f}}

\newcommand{\CF}{C_F}
\newcommand{\TR}{T_R}
\newcommand{\CA}{C_A}

\newcommand{\nf}{n_{\!f}}

\long\def\symbolfootnote[#1]#2{\begingroup%
\def\thefootnote{\fnsymbol{footnote}}\footnote[#1]{#2}\endgroup} 
\begin{document}

% $Id: autolet-plb.tex,v 1.15 2004/01/15 11:16:05 salam Exp $

%----------------------------------------------------------------

\begin{titlepage}
\renewcommand{\thefootnote}{\fnsymbol{footnote}}
\begin{flushright}
     Bicocca-FT-03-8 \\
     DCPT-03-36\\
     IPPP-03-18\\
     LPTHE-03-12\\
     NIKHEF/2003-007\\
\end{flushright}
\par \vskip 10mm
\begin{center}
  {\Large \bf Generalized resummation of QCD final-state observables}
\end{center}
\par \vskip 2mm
\begin{center}
{\bf 
A. Banfi$^{(1)}$, G.P. Salam$^{(2)}$ and
G. Zanderighi$^{(3)}$\symbolfootnote[1]{Current address: Fermi
  National Accelerator Laboratory, Batavia, IL 60510-500, USA.}}\\[4mm]
\begin{minipage}{6.5in}
$^{(1)}$ NIKHEF Theory Group, P.O. Box 41882, 1009 DB Amsterdam,
  The Netherlands, \\
Dipartimento di Fisica, Universit\`a di 
  Milano-Bicocca and INFN,
  Sezione di Milano, Italy. \\[2mm]
$^{(2)}$ LPTHE, Universities of Paris VI and VII and CNRS UMR 7589, 
 Paris, France. \\[2mm]
$^{(3)}$  IPPP,
  Department of Physics, University of Durham, Durham DH1 3LE,
  UK.
\end{minipage}
\end{center}

\par \vskip 2mm
\begin{center} {\large \bf Abstract} \\

\end{center}
\begin{quote}
  The resummation of logarithmically-enhanced terms to all
  perturbative orders is a prerequisite for many studies of QCD
  final-states.  Until now such resummations have always been
  performed by hand, for a single observable at a time. In this letter
  we present a general `master' resummation formula (and applicability
  conditions), suitable for a large class of observables. This makes
  it possible for next-to-leading logarithmic resummations to be
  carried out automatically given only a computer routine for the
  observable.  To illustrate the method we present the first
  next-to-leading logarithmic resummed prediction for an event shape in
  hadronic dijet production.
\end{quote}
\end{titlepage}

%----------------------------------------------------------------

\section{Introduction \label{sec:Int}}
QCD is unique among the theories of the standard model in that both
strong and weak coupling regimes are relevant to modern collider
experiments. This manifests itself most dramatically in hadronic final
states of high-energy collisions, whose branching pattern is sensitive
to physics spanning the whole range of scales from the (perturbative)
hard collision virtuality down to (non-perturbative) hadronic masses.
Accordingly final states are a privileged laboratory for QCD studies:
perturbative investigations have for example led to many measurements
of the strong coupling, $\as$ \cite{Bethke}, and to tests of the
underlying $\mathrm{SU}(3)$ group structure of the theory \cite{SU3};
and final-states are also proving to be a rich source of information
on the poorly understood relation between perturbative, partonic
predictions and the non-perturbative, hadronic degrees of freedom
observed in practice \cite{MLLA,DMW}.

Among the most widely studied final-state properties are measures
($v$) of the extent to which the geometric properties of an event's
energy-momentum flow differ from that of a Born event (the lowest
order contribution to the given process, for example $\ee\to q\bar q$).
Fixed-order 
perturbative calculations, which involve a small number of additional
partons, are suitable for describing large departures from the
Born-event energy flow pattern, in which the extra partons are
energetic and at large angles.  Such configurations are however rare,
their likelihood being suppressed by powers of the
perturbative coupling.

The most common events are instead those in which the departure from
the Born energy-flow pattern is small, $v\ll 1$, with any extra partons
being soft and/or collinear to the original Born-event partons. This
poses a problem for fixed-order studies because each power of the
coupling is then accompanied by up to two powers of the large
logarithm $\ln 1/v$, associated with soft and collinear divergences.
As a result, the perturbative series involves terms $(\as \ln^2
1/v)^n$, and must be resummed to all orders.

Today's state of the art calculations exploit the fact that for many
measures (`observables'), the dominant all-orders perturbative
contribution can be written as an exponential of leading-logarithmic
(LL) terms $\as^n\ln^{n+1} 1/v$. Furthermore the next-to-leading
logarithmic (NLL) terms, $\as^n\ln^{n} 1/v$, factorise and can be
calculated to all orders \cite{CTTW}.
But to obtain this NLL accuracy one needs a detailed understanding of
the observable's analytical properties and of the corresponding
phase-space integrals. Thus it is usual for an entire paper to be
dedicated to the resummation, in a single process, of just one or two
observables.

In this letter we instead adopt the novel approach of simultaneously
examining a whole class of observables, for which it will be possible
to carry out a common analysis. The
results, involving a `master' formula with applicability conditions,
will be relevant to a range of processes including $\ee$ to $2$ or $3$
jets, DIS to $1$ or $2$ jets, Drell-Yan (or $\gamma$, $W^{\pm}$,
Higgs,\ldots) plus a jet, and hadronic dijet production.
The final answer for some specific
observable will be expressed in terms of straightforwardly (and
automatically) identifiable characteristics of the observable. 

% The purpose of this letter is to present a novel approach to
% final-state resummation, based on a `master' formula appropriate for a
% large class of observables in a range of processes, including $\ee$ to
% $2$ or $3$ jets, DIS to $1$ or $2$ jets, Drell-Yan (or $\gamma$,
% $W^{\pm}$, Higgs,\ldots) plus a jet, and hadronic dijet production.
% % 
% The master formula is accompanied by a set of conditions that must be
% satisfied by the observable in order for the approach to be valid.

%----------------------------------------------------------------
\section{Master formula and applicability conditions \label{sec:mast}}
Let us start by taking a Born event consisting of $n$ hard partons or
`legs' ($n_i$ of which are incoming), with momenta $p_1,\ldots,p_n$. We
shall consider the resummation, in the $n$-jet limit, of $n$-jet
infrared and collinear (IRC) safe observables --- these measure the
extent to which an event's energy flow departs from that of an
$n$-parton event. For the resummation approach to be valid the
observable (a function $V$ of the final-state momenta) should:
\begin{enumerate}
\item vanish smoothly as a single extra ($n$+$1$)$^\mathrm{th}$ parton
  of momentum $k$ is made asymptotically soft and collinear to leg $\ell$, the
  functional dependence being of the form:
  \begin{equation}
    \label{eq:simple}
    V(\{{\tilde p}\}, k)=
    d_{\ell}\left(\frac{k_t}{Q}\right)^{a_\ell}e^{-b_\ell\eta}\, 
    g_\ell(\phi)\>.
  \end{equation}
  Here $Q$ is a hard scale of the problem; $\{{\tilde p}\}$ represents
  the Born (hard) momenta after recoil from the emission, which is defined in
  terms of its transverse momentum $k_t$ and rapidity $\eta$ with
  respect to leg $\ell$, and where relevant, by an azimuthal angle
  $\phi$ relative to a Born event plane. By requiring the functional
  form \eqref{eq:simple} (in practice, almost always valid), the
  problem of analysing the observable reduces in part to identifying,
  for each leg $\ell$, the coefficients $a_\ell$, $b_\ell$, $d_\ell$
  as well as the function $g_\ell(\phi)$ parameterising the azimuthal
  dependence (the normalisation may be fixed by the condition
  $g_\ell(\pi/2)=1$). IRC safety implies $a_\ell > 0$ and $b_\ell >
  -a_\ell$ (see also \cite{BKS03}). We further require the observable
  to be positive definite.
  
\item be \emph{recursively} IRC (rIRC) safe: meaning that, given an
  ensemble of arbitrarily soft and collinear emissions, the addition
  of a relatively much softer or more collinear emission should not
  significantly alter the value of the observable.  The formal
  requirement can be formulated as follows: we introduce momenta
  $\kappa_i(\lambda_i)$ that are \emph{functions} of parameters
  $\lambda_i$ such that,
  \begin{equation}
    \label{eq:kidef}
    V(\{\tilde p\},\kappa_i(\lambda_i)) = \lambda_i \,,
  \end{equation}
  with the condition that in the soft and/or collinear limits, $\lambda_i
  \to 0$, the azimuthal angle $\phi_i$ of $\kappa_i(\lambda_i)$ should be
  fixed.  Each of the momentum functions $\kappa_1(\lambda)$,
  $\kappa_2(\lambda)$, etc.  may be different as long as they all satisfy
  eq.~(\ref{eq:kidef}).  The conditions for rIRC safety then become that
  \begin{itemize}
  \item[(a)] the limit 
    \begin{equation}
      \label{eq:rIRClimit1}
      \lim_{\epsilon \to 0} \frac{1}{\epsilon} 
      V(\{\tilde p\},\kappa_1(\epsilon \lambda_1), \ldots,
      \kappa_m(\epsilon \lambda_m))
    \end{equation}
    should be well-defined and non-zero (except possibly in a region
    of phase-space of zero measure). This can be interpreted as a
    requirement that the soft and collinear
    scaling properties of the observable should be the same regardless
    of whether there is just one, or many emissions.
  \item[(b)] the following two limits should be identical,
    \begin{multline}
      \label{eq:rIRClimit2}
      \lim_{\lambda_{m+1}\to 0} \lim_{\epsilon \to 0} \frac{1}{\epsilon} 
      V(\{\tilde p\},\kappa_1(\epsilon \lambda_1), \ldots, \kappa_m(\epsilon
      \lambda_m), \kappa_{m+1}(\epsilon
      \lambda_{m+1})) \\= 
      \lim_{\epsilon \to 0} \frac{1}{\epsilon} 
      V(\{\tilde p\},\kappa_1(\epsilon \lambda_1), \ldots, \kappa_m(\epsilon
      \lambda_m))\,, 
    \end{multline}
    \ie\ having taken the limit eq.~(\ref{eq:rIRClimit1}), the
    addition of an extra much softer and/or more collinear emission
    should not affect the value of the of the observable. At first
    sight this closely resembles normal IRC safety, but actually differs
    critically because of the order of the limits on the left-hand
    side of eq.~(\ref{eq:rIRClimit2}).
  \end{itemize}
  These conditions (until now never formulated), which should hold
  regardless of how precisely the $\kappa_i(\lambda)$ vanish as
  $\lambda \to 0$, allow one to translate a restriction on the
  ensemble of emissions, $V(\{\tilde p\},k_1,\ldots,k_m) < v$, into a
  restriction on each individual emission, $V(\{\tilde p\},k_i)
  \lesssim v$ (modulo NLL corrections discussed below). This is
  necessary in order to ensure exponentiation of the LL
  terms.\footnote{An interesting exercise is to verify that the JADE
    3-jet resolution parameter in $\ee$, which is known not to
    exponentiate \cite{JadeDL}, is indeed rIRC unsafe.}

\item be continuously global \cite{NG} --- this means that for a
  single soft emission, the observable's parametric dependence on the
  emission's transverse momentum (with respect to the nearest leg)
  should be independent of the emission direction, $\partial_\eta
  \partial_{\ln k_t} \ln V(\{{\tilde p}\}, k) = 0$ and $\partial_\phi
  \partial_{\ln k_t} \ln V(\{{\tilde p}\}, k) = 0$. In practice, this
  is perhaps the most restrictive of the conditions. It avoids the
  need to analyse possibly quite complicated angular boundaries
  between regions with different transverse-momentum dependences and
  calculate the corresponding non-global logarithms~\cite{NG}. It implies $a_1 =
  a_2 = \ldots = a_n \equiv a$.
\end{enumerate}

Given the above conditions, one can derive the
following NLL master resummation formula for the probability
$\Sigma(v)$ that the observable's value is less than $v$
\cite{BSZPrep}:
\begin{equation}
\begin{split}
  \label{eq:Master}
  \ln \Sigma(v) &= -\sum_{\ell=1}^n C_\ell \left[r_\ell(L) + 
    r_\ell'(L) \left(\ln {\bar d}_\ell - b_\ell \ln
      \frac{2E_\ell}{Q}\right) 
    + B_\ell \, T\!\left(\frac{L}{a+b_\ell}\right)
  \right] \\
  &+ \sum_{\ell=1}^{n_i} \ln \frac{f_\ell(x_\ell,v^{\frac{2}{a+b_\ell}}
    \muf^2)}{f_\ell(x_\ell, \muf^2)}  
  + \ln S\left(T(L/a\right)) 
  + \ln \cF(C_1 r_1',\ldots,C_n r_n')
  \,,
\end{split}
\end{equation}
%\begin{multline}
%  \label{eq:Master}
%  \ln \Sigma(v) = -\sum_{\ell=1}^n C_\ell \left[r_\ell(L) + 
%    r_\ell'(L) \left(\ln {\bar d}_\ell - b_\ell \ln
%      \frac{2E_\ell}{Q}\right)
%    \right. \\ \left . 
%    + B_\ell \, T\!\left(\frac{L}{a+b_\ell}\right)
%  \right] %\\
%  + \sum_{\ell=1}^{n_i} \ln \frac{f_\ell(x_\ell,v^{\frac{2}{a+b_\ell}}
%    \muf^2)}{f_\ell(x_\ell, \muf^2)} 
%  \\  
%  + \ln S\left(T(L/a\right)) 
%  + \ln \cF(C_1 r_1',\ldots,C_n r_n')
%  \,,
%\end{multline}
where $L = \ln 1/v$, $C_\ell$ is the colour factor associated with Born
leg $\ell$ ($C_F$ for a quark and $C_A$ for a gluon), and $E_\ell$ is
its energy, $B_\ell$ accounts for hard collinear splittings and is
$-3/4$ for quarks and $-(11\CA - 4\TR \nf)/(12\CA)$ for gluons, $\ln
{\bar d}_\ell = \ln d_\ell + \int_0^{2\pi} \frac{d\phi}{2\pi} \ln
g_\ell(\phi)$, and for incoming legs, the $f_\ell$ are the appropriate
(Born flavour) parton densities.
%
% MOVED BELOW
%We note that: eq.~\eqref{eq:Master} is independent of the frame in
%which one determines the $d_\ell$, because the frame-dependence of the
%$d_\ell$ is cancelled by that of the $E_\ell$; to NLL accuracy,
%\eqref{eq:Master} is also independent of the choice of hard scale $Q$;
%finally the continuous globalness of the observable ensures that, to
%NLL accuracy, eq.~\eqref{eq:Master} is insensitive to the details of
%the observable's dependence on large-angle soft gluons.

The functions $r_\ell(L)$ contain all the LL (and some NLL) terms and
are defined by
\begin{equation}
  r_\ell(L) =
  \int_{Q^2 v^{\frac2a}}^{Q^2
    v^{\frac{2}{a+b_\ell}}
  }\frac{dk_t^2}{k_t^2}\frac{\as(k_t)}{\pi}
  \ln\left(\frac{k_t}{v^{1/a}Q}\right)^{a/b_\ell}
  %\\
  + \int_{Q^2
    v^{\frac{2}{a+b_\ell}}
  }^{Q^2} \frac{dk_t^2}{k_t^2}\frac{\as(k_t)}{\pi}\ln\frac{Q}{k_t}\>,
\end{equation}
%\begin{multline}
%  \label{eq:rad-dl}
%  r_\ell(L) =
%  \int_{Q^2 v^{\frac2a}}^{Q^2
%    v^{\frac{2}{a+b_\ell}}
%  }\frac{dk_t^2}{k_t^2}\frac{\as(k_t)}{\pi}
%  \ln\left(\frac{k_t}{v^{1/a}Q}\right)^{a/b_\ell}
%  +\\
%  + \int_{Q^2
%    v^{\frac{2}{a+b_\ell}}
%  }^{Q^2} \frac{dk_t^2}{k_t^2}\frac{\as(k_t)}{\pi}\ln\frac{Q}{k_t}\>,
%\end{multline}
where $\as$ runs at two-loop order and is to be taken in the
Bremsstrahlung scheme \cite{CMW}. 
Exponentiation guarantees that the
LL terms of $r_\ell$ are in the class $\as^n L^{n+1}$.

All remaining terms are relevant only at NLL accuracy: $r_\ell' =
\partial_L r_\ell$; $T(L)$ is given by
\begin{equation}
  \label{eq:T}
  T(L) = \int_{Q^2 e^{-2L}
}^{Q^2}\frac{dk_t^2}{k_t^2}\frac{\as(k_t)}{\pi}\>.
\end{equation}
The process dependence associated with large-angle soft radiation is
contained in $S(T(L/a))$, whose form depends on the number of legs:
\begin{align*}
  n = 2:\!\! &\quad \ln S(t) = -t\cdot 2C_F \,\ln \frac{Q_{qq'}}{Q}\,,\\
  n = 3:\!\! &\quad \ln S(t) = -t \left[{\CA}\ln \frac{Q_{qg} Q_{q'
        g}}{Q_{q q'} Q}  
+ 2\CF \ln \frac{Q_{q q'}}{Q}\right] \,,\\
  n = 4:\!\! &\quad \ln S(t) = -t \sum_\ell C_\ell \ln \frac{Q_{12}}{Q}
     %\\ & \qquad \qquad\qquad\quad
       + \ln \frac{\mathrm{Tr} (H
       e^{-t\Gamma^\dagger/2} M
       e^{{-t}\Gamma/2})}{\mathrm{Tr} (H M)} \,,
\end{align*}
where $Q^2_{ab} = 2p_a . p_b$ and $q$, $q'$ and $g$ denote the
(anti)-quarks and gluon.  The $n=2,3$ formulae apply to $\ee$, DIS and
Drell-Yan production, while a process such as $gg\to \mathrm{Higgs}
+g$ would simply involve different colour factors.  The $n=4$ formula
applies to hadronic dijet production ($1$ and $2$ label the
incoming legs).  The quantities $H$, $M$ and $\Gamma$ are the hard,
soft and anomalous dimension matrices of \cite{Sterman4Legs} (modulo
normalisations and our explicit extraction of the factor $t$ from
$\Gamma$, see \cite{BSZPrep}).

Finally, we examine the factor $\cF$.  Without it,
eq.~\eqref{eq:Master} corresponds essentially to the probability of
vetoing all (independent) emissions $k$ with $V(\{\tilde p\},k) > v$.
But given some ensemble of emissions $k_1,\ldots, k_m$
that individually satisfy $V(\{\tilde p\},k_i) < v$, the observable
may be such that one still has $V(\{\tilde p\},k_1,\ldots, k_m) > v$.
It is then necessary to apply a somewhat stronger veto in order to
guarantee $V(\{\tilde p\},k_1,\ldots, k_m) < v$.  This (and the
converse situation of $V(\{\tilde p\},k_i) > v$ being allowed in the
presence of multiple emissions) is accounted for by the NLL function
$\cF$,
%\begin{multline}
\begin{equation}
  \label{eq:cF}
  \cF(R_1',\ldots, R_n') = 
  \left\langle
    \exp \left\{ 
      -R' \ln 
    \frac{V(\{{\tilde p}\},k_{1},\ldots, k_m) }{\max \{V(\{{\tilde
        p}\},k_1),\ldots, V(\{{\tilde p}\},k_m)\}} \right\}
  \right \rangle,
\end{equation}
%\end{multline}
where $R' = \sum_\ell R_\ell'$, $ R_\ell'= C_\ell\, r'_\ell$.  The
average is carried out over ensembles of emissions generated as
follows (\cnf section 2 of \cite{BSZ}):
first one specifies the value of the maximum of the $V(\{{\tilde
  p}\},k_i)$, say $v_{\max}$ ($\ll1$). For each event (ensemble), a random
number ($m$, formally infinite) of emissions is generated, according
to an independent emission pattern uniform in $\ln k_t$, $\eta$ and
$\phi$, such that on average, below $v_{\max}$, the density per unit
$\ln V(\{\tilde p\},k)$ of emissions on leg $\ell$ is $R'_\ell$.
% All emissions are then
%distributed according to an independent emission pattern uniform in
%$\ln k_t$, $\eta$ and $\phi$ such that the density in $\ln V(\{p\},k)$
%of emissions along leg $\ell$ is $R_\ell'$. 
To ensure a result
containing only NLL terms, one takes the result in the limit
$v_{\max}\to0$.  Full details, including the derivation and a
treatment of subtleties associated with the running of the coupling
and the recoil momenta, $\{\tilde p\}$ (determined anew for each set
of emitted momenta), are given elsewhere \cite{BSZ,BSZPrep}.  

We note that attempting to evaluate $\cF$ for an observable that is
rIRC unsafe will yield a result that is either ill-defined or
improperly behaved for $R'\to 0$. This can be thought of as analogous
to the divergence of NLO terms of a fixed order calculation for
observables that are IRC unsafe.

% HERE? 
%We note that: 
Before proceeding, some remarks on the master formula are in order. 
As can be verified in a straightforward way, we note that: 
\begin{itemize}
\item eq.~\eqref{eq:Master} is independent of the frame in which one
  determines the $d_\ell$, because the frame-dependence of the
  $d_\ell$ is cancelled by that of the $E_\ell$;
\item to NLL accuracy, eq.  \eqref{eq:Master} is also independent of
  the choice of hard scale $Q$;
\item hard emissions collinear to each leg $\ell$ are accounted for
  through the factor $B_\ell$ in eq.~\eqref{eq:Master}, and, in the case
  of radiation from an incoming leg, also through the modification of the
  corresponding parton density factorisation scale from $\muf$ to
  $\muf v^{\frac{1}{a+b_\ell}}$.  Hard collinear contributions depend
  only on the combination $a+b_\ell$, which is to be related to the
  fact that in this region, for an emission $k$ with a fixed energy
  fraction, the observable behaves simply as 
  $V(\{\tilde p\},k)\sim (k_t/Q)^{a+b_\ell}$;
  
\item finally the continuous globalness of the observable ensures
  that, to NLL accuracy, eq.~\eqref{eq:Master} is insensitive to the
  details of the observable's dependence on large-angle soft gluons,
  the only relevant information being that for any large-angle
  emission the observables scales as $V(\{\tilde p\},k)\sim
  (k_t/Q)^a$.
\end{itemize}

%ADDED BACK 
Given the above elements, one could imagine a procedure whereby the
applicability conditions and the parameters of eq.~\eqref{eq:simple}
are established by hand, analytically, with only the $\cF$ being
determined numerically. A related approach was presented in
\cite{BSZ}, though instead of using a master formula, we had to
analytically carry out a resummation for a `simplified' version of the
full observable --- new results were obtained there for three
observables in $\ee\to$~2~jets.  This was already a considerable
improvement over the traditional, entirely manual resummation
approach, which requires a painstaking analysis of the observable's
dependence on arbitrary numbers of emissions followed by involved
mathematical procedures to obtain a result which quite often cannot
even be expressed in closed form (see \cite{eeKout} for a tortuous
example).

However the introduction of a master formula makes it possible to
implement a fundamentally new approach. Given a subroutine that
calculates the observable for an arbitrary set of four momenta, a
computer program can carry out the entire resummation: it first
establishes whether the applicability conditions hold true and
determines for each leg $\ell$ the parameters and functions of
\eqref{eq:simple}, $a_\ell$, $b_\ell$, $d_\ell$ and
$g_\ell(\phi)$.\footnote{In our current implementation, for technical
  reasons, $a_\ell$ and $b_\ell$ are restricted be multiples of $1/4$,
  but the extension to any power of $1/2$ is trivial.}
% GZ: the following is not correct!
%  $b_\ell$ have to be multiples of $1/4$, but the extension to a wider
%  range of {\it rational} numbers is in principle straightforward.}
This is achieved by probing the observable with randomly chosen test
configurations of soft and collinear emissions, taking the asymptotic
limit with the help of high precision arithmetic (we choose to use
Bailey's portable multiple-precision package \cite{MP}).

If any of the applicability conditions fail to hold (\eg for the Jade
3-jet resolution parameter in $\ee$, which is not recursively IRC safe
and so does not exponentiate \cite{JadeDL}), the program does not
proceed, \ie a resummed answer is provided only when the correctness
of the result is guaranteed to NLL accuracy.

This method allows one to make an attempt at the resummation of an
arbitrary observable in a fully automated way, accessible
even to non-experts.  This is to be compared to the standard
approach, involving a painstaking (and historically sometimes
error-prone) manual analysis of the observable, requiring a search for
integral transformations (in up to 5 variables \cite{eeKout}!) to
reduce it to a factorised form ---  this form is then used for the actual
resummation, after which one evaluates the inverse transforms.
%
%With the above elements, the resummation of an arbitrary observable
%becomes straightforward --- given a subroutine for calculating the
%observable, a computer program can verify the applicability conditions
%and determine the coefficients $a_\ell$, $b_\ell$ and $d_\ell$ and the
%functions $g_\ell(\phi)$ and $\cF(R')$, thus providing all the
%elements needed for the resummation (in the practical implementation
%we made extensive use of Bailey's portable
%multiple-precision package \cite{MP}). Such a procedure is to be
%compared to the 
%standard approach, involving a painstaking (and historically sometimes
%error-prone) manual analysis of the observable, requiring a search for
%integral transformations (in up to 5 variables \cite{eeKout}!) to
%reduce it to a factorized form. This form is used for the actual
%resummation, after which one evaluates the inverse transforms.

Usually the two approaches give indistinguishable results, though in
some instances one or the other may be preferred: for some
observables (involving cancellations between contributions from
different emissions),
exponentiation is only partial, resulting in
\eqref{eq:Master} being accurate only up to some finite value of 
%$R'\sim 1$ 
$R'$ (typically of order 1)
--- beyond this point $\cF$ diverges \cite{BSZ} and only the
use of the 
appropriate integral transform method can give a full answer (\eg
Drell-Yan $p_t$ resummations with a Fourier transform to impact
parameter). For certain other observables however, the `factorising'
integral transform has yet to be found (\eg the Durham 3-jet resolution
parameter) and a numerical approach represents the only way of
obtaining a resummed answer.

%----------------------------------------------------------------
\section{Resummation in hadronic dijets events \label{sec:hh}}
We have verified that our approach reproduces the analytically known
results in $\ee$ and DIS (\eg \cite{CTTW,eeKout,NG}).  Here, to
demonstrate its feasibility more generally, we show the first resummed
result for an event shape in hadronic dijet production. Rapid progress
is currently being made on measurements \cite{D0} and fixed-order
predictions \cite{NLOJET} for such observables, with the results
showing a clear need for resummations. We shall examine the (global)
transverse thrust (as opposed to D\O's discontinuously global variant
\cite{D0}), defined as:
\begin{equation}
  \label{eq:Ttg}
  T_{\perp} \equiv \max_{\vec n_\perp} \frac{\sum_i |{\vec
      p}_{\perp i}\cdot {\vec
      n_\perp}|}{\sum_i p_{\perp i}}\,,
\end{equation}
where the sum runs over all particles in the final state, $p_{\perp}$
is the momentum transverse to the beam direction (rather than to a
given leg, denoted by $p_t$) and $\vec n_\perp$ is the unit transverse
vector that maximises the projection.

The transverse thrust has a couple of features worth commenting:
firstly, it receives non-negligible contributions from emissions
nearly collinear to the beams --- thus it will be sensitive to
radiation from the `underlying event', making it useful for
quantitative studies of non-perturbative effects that are
qualitatively new compared to those examined up to now in $\ee$ and
DIS.  Various other observables will be proposed in forthcoming work
\cite{BSZPrep}, a number of which will be less sensitive to radiation
from the incoming legs, providing a good degree of complementarity.
Secondly, whereas \eqref{eq:Ttg} sums over all particles, experiments
can only measure up to some maximum rapidity $\eta_{\max}$. In the
presence of such a restriction it can be shown that the resummation
still remains valid for values of $v \gtrsim
e^{-(a+b_{\min})\eta_{\mathrm{max}}}$, where $b_{\min}$ is the smaller
of the two incoming leg $b_\ell$ values \cite{BSZPrep}.

\begin{table}%[H] add [H] placement to break table across pages
\begin{center}
 \begin{tabular}{| c | c | c | c | c | c |}
 \hline
 leg $\ell$ & $a_{\ell}$ & $b_{\ell}$ & $g_{\ell}(\phi)$ & $d_{\ell}$ & $ \langle \ln g_{\ell}(\phi) \rangle$  \\
 \hline
 \hline
1 & 1 & $ 0$ & tabulated &  1.02062 & $ -1.85939$ \\ 
 \hline
2 & 1 & $ 0$ & tabulated &  1.02062 & $ -1.85939$ \\ 
 \hline
3 & 1 & $ 1$ &$\sin^2\phi$ &  1.04167 & $-2\ln(2)$ \\ 
 \hline
4 & 1 & $ 1$ &$\sin^2\phi$ &  1.04167 & $-2\ln(2)$ \\ 
 \hline
 \end{tabular}
\caption{Automatically determined leg parameters for $\tau_{\perp}$ 
  in hadronic dijet production (in a c.o.m.\ frame
  with outgoing legs at an angle $\cos \theta = 0.2$).\label{tab:legsothrhh}}
%\begin{ruledtabular}
%\end{ruledtabular}
\end{center}\end{table}

Let us now examine the automated resummation itself: the quantity to
be resummed is actually $\tau_{\perp}\equiv 1-T_{\perp}$, since it is
this that vanishes in the Born limit. The observable passes all the
(automated) applicability tests and table~\ref{tab:legsothrhh},
generated automatically,
shows the leg properties for a particular reference Born
configuration.  
The different $b_\ell$ values for
incoming and outgoing legs imply different leading logarithmic
structures.  The azimuthal dependence $g_\ell(\phi)$ is tabulated and
integrated numerically, except in the case of certain easily
recognisable analytical functions.
The function $\cF$ has a simple analytical form,\footnote{It can be
   automatically established that $\tau_\perp$ is additive, $V(\{\tilde
   p\}, k_1,\ldots, k_m) = \sum_i V(\{\tilde p\}, k_i)$, implying $\cF = 
   e^{-\gamma_E R'}/\Gamma(1+R')$ \cite{CTTW}.}
%\cite{cFAnl},
however to demonstrate the feasibility of our whole approach we shall
show results based on a numerically determined~$\cF$.
%The function $\cF$ has a simple analytical
%form \cite{cFAnl}, however to demonstrate the feasibility of our whole
%approach we shall show results based on a numerically
%determined~$\cF$.

%It so happens (also established automatically) that $\tau_{\perp}$
%belongs to the special class of \emph{additive} observables, those
%satisfying $V(\{{\tilde p}\}, k_1,\ldots, k_m) = V(\{{\tilde p}\},
%k_1) + \ldots + V(\{{\tilde p}\},k_m)$. For such (relatively common)
%observables, $\cF$ is known analytically, $e^{-\gamma_E
%  R'}/\Gamma(1+R')$ \cite{CTTW}, and the program would normally make
%use of this information. However so as to demonstrate the feasibility
%of our whole approach, we shall show results based on a numerically
%determined~$\cF$.

One further step is needed before presenting actual distributions: our
master formula applies to individual Born configurations. For example
in the case of $p\bar p \to $~2~jets, the analysis is carried out with
a fixed rapidity for the pair of jets and fixed values of the
Mandelstam invariants of the underlying hard process. In contrast
experimental measurements integrate over a range of Born
configurations. A priori there is no reason for the leg parameters or
$\cF$ to be independent of the configuration and it could be necessary
to repeat the analysis for a range of configurations. However for most
observables, modulo certain permutations of momenta (as can %once again
be verified automatically), it is only the $d_\ell$ that depend on the
configuration and they are easily redetermined as one integrates over
Born configurations.

\begin{figure}
\begin{center}
\includegraphics[width=0.48\textwidth]{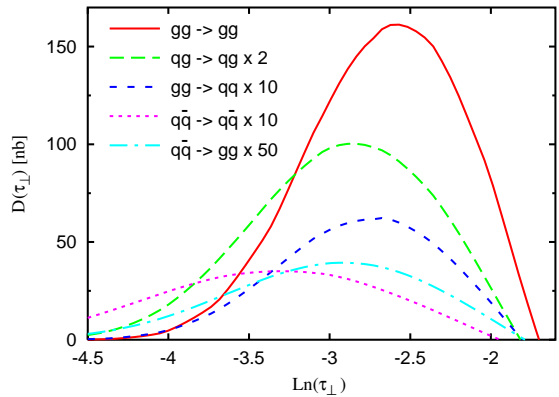}%
\end{center}
\caption{NLL resummed differential distribution $D(\tau_\perp) \equiv
  d\Sigma(\tau_\perp)/d \ln \tau_\perp$ for different underlying hard
  subprocesses.
  \label{fig:othrDist}}
\end{figure}

The resulting distribution for $\tau_{\perp}$ is shown in
fig.~\ref{fig:othrDist}, decomposed into the most relevant underlying
hard subprocesses, for the Tevatron run II regime ($\sqrt{s} =
1.96$TeV).  We select events containing two outgoing jets with
$E_\perp > 50$GeV and $|\eta|<1.0$ and use the CTEQ6M parton density
set~\cite{CTEQ}, corresponding to $\alpha_s(M_Z) = 0.118$.  We have
set $Q=\mu_F = \mu_R$ to be the Born partonic c.o.m. energy, though in
future work we intend to explore a range of alternative scales.  As is
to be expected, channels with lower overall colour charge have broader
distributions. We note that the different shapes of the various
channels constitutes information that might be exploitable in fits of
parton distributions.  Of course detailed phenomenological analyses,
both for perturbative and non-perturbative quantities, will also
require matching to fixed-order predictions, another step that we
leave to future work.
Here we just remark that resummed results obtained from the master
formula are in semi-analytical form (fully analytic but for the pure
NLL function $\cF$), so that they can be easily expanded to give the
fixed-order coefficients needed when matching.

%----------------------------------------------------------------
\section{Conclusions \label{sec:concl}}
In this letter we have provided the elements needed for a novel,
automated approach to general NLL resummation, specifically for the
case of continuously global, exponentiable $(n\!+\!1)$-jet final-state
observables in the $n$-jet limit. Results are obtained simply by
specifying the Born process (and the number of hard partons) and
providing the definition of the observable to be resummed in the form
of a computer routine, similar to the long-established practice for
fixed-order calculations, and in contrast to the tedious manual
approach that has been used up to now for resummations. The results
are provided in semi-analytical form, making it straightforward to
obtain the expansions needed for procedures such as matching to
fixed-order predictions.

We have demonstrated that the approach can be implemented in practice,
by presenting automatically generated predictions for the transverse
thrust in hadronic dijet production, the first event shape to be
resummed in this important process. Only concerns for brevity prevent
us from showing results for a range of other observables and
processes, including several new observables in hadronic dijet
production and jet rates in $\ee$ and DIS.

An open question is whether such an approach, based on the analysis of
\emph{classes} of observables can be applied in other resummations
contexts, or in the search for higher resummation accuracies. We
enthusiastically advocate investigations in this direction.

%----------------------------------------------------------------
\paragraph{Acknowledgments}
We wish to thank Mrinal Dasgupta, Yuri Dokshitzer, Eric Laenen and
Pino Marchesini for useful discussions and suggestions and Zoltan
Nagy for providing us with the latest version of NLOJET++ and
assistance in using it.  We are grateful to each other's institutes
for hospitality and to CERN and the University of Milano-Bicocca for
the use of computing facilities.

%----------------------------------------------------------------


\begin{thebibliography}{99}
\bibitem{Bethke} 
S.~Bethke,
%``Determination of the QCD coupling alpha(s),''
J.\ Phys.\ G {\bf 26}, R27 (2000).
%[arXiv:hep-ex/0004021].
%%CITATION = HEP-EX 0004021;%%

\bibitem{SU3}
S.~Kluth et al.,
%``A measurement of the QCD colour factors using event shape distributions  at s**(1/2) = 14-GeV to 189-GeV,''
Eur.\ Phys.\ J.\ C {\bf 21}, 199 (2001) and references therein.
%[arXiv:hep-ex/0012044].
%%CITATION = HEP-EX 0012044;%%


\bibitem{MLLA} 
V.~A.~Khoze and W.~Ochs,
%``Perturbative QCD approach to multiparticle production,''
Int.\ J.\ Mod.\ Phys.\ A {\bf 12}, 2949 (1997) and references therein.
%[arXiv:hep-ph/9701421].
%%CITATION = HEP-PH 9701421;%%

\bibitem{DMW} Yu.~L.~Dokshitzer, G.~Marchesini and B.~R.~Webber,
%``Dispersive Approach to Power-Behaved Contributions in QCD Hard Processes,''
Nucl.\ Phys.\ B {\bf 469}, 93 (1996);
%[arXiv:hep-ph/9512336].
%%CITATION = HEP-PH 9512336;%%
see also M.~Beneke,
%``Renormalons,''
Phys.\ Rept.\  {\bf 317}, 1 (1999).
%[arXiv:hep-ph/9807443].
%%CITATION = HEP-PH 9807443;%%

\bibitem{CTTW}
S.~Catani, L.~Trentadue, G.~Turnock and B.~R.~Webber,
%``Resummation of large logarithms in e+ e- event shape distributions,''
Nucl.\ Phys.\ B {\bf 407}, 3 (1993).
%%CITATION = NUPHA,B407,3;%%

\bibitem{BKS03} 
C.~F.~Berger, T.~Kucs and G.~Sterman,
%``Event shape / energy flow correlations,''
Phys.\ Rev.\ D {\bf 68}, 014012 (2003)
%hep-ph/0303051.
%%CITATION = HEP-PH 0303051;%%



\bibitem{JadeDL}
N.~Brown and W.~J.~Stirling,
%``Jet Cross-Sections At Leading Double Logarithm In E+ E- Annihilation,''
Phys.\ Lett.\ B {\bf 252}, 657  (1990).%\\
%%CITATION = PHLTA,B252,657;%%

\bibitem{NG} 
M.~Dasgupta and G.~P.~Salam,
%``Resummation of non-global QCD observables,''
Phys.\ Lett.\ B {\bf 512}, 323 (2001);
%[arXiv:hep-ph/0104277].
%%CITATION = HEP-PH 0104277;%%
%M.~Dasgupta and G.~P.~Salam,
%``Resummed event-shape variables in DIS,''
JHEP {\bf 0208}, 032 (2002).
%[arXiv:hep-ph/0208073].
%%CITATION = HEP-PH 0208073;%%


\bibitem{BSZPrep} A.~Banfi, G.~P.~Salam and G.~Zanderighi, in
  preparation. 
  

\bibitem{CMW} 
S.~Catani, B.~R.~Webber and G.~Marchesini,
%``QCD coherent branching and semiinclusive processes at large x,''
Nucl.\ Phys.\ B {\bf 349}, 635 (1991);
%%CITATION = NUPHA,B349,635;%%
Yu.~L.~Dokshitzer, V.~A.~Khoze and S.~I.~Troyan,
%``Specific features of heavy quark production. LPHD approach to heavy particle spectra,''
Phys.\ Rev.\ D {\bf 53}, 89 (1996).
%[hep-ph/9506425].
%%CITATION = HEP-PH 9506425;%%

\bibitem{Sterman4Legs}
J.~Botts and G.~Sterman,
%``Hard Elastic Scattering In QCD: Leading Behavior,''
Nucl.\ Phys.\ B {\bf 325}, 62 (1989);
%%CITATION = NUPHA,B325,62;%%
%;\\
N.~Kidonakis, G.~Oderda and G.~Sterman,
%``Evolution of color exchange in {QCD} hard scattering,''
Nucl.\ Phys.\ B {\bf 531}, 365 (1998).
%[arXiv:hep-ph/9803241]
%%CITATION = HEP-PH 9803241;%%

\bibitem{BSZ} 
A.~Banfi, G.~P.~Salam and G.~Zanderighi,
%``Semi-numerical resummation of event shapes,''
JHEP {\bf 0201}, 018 (2002).
%[arXiv:hep-ph/0112156].
%%CITATION = HEP-PH 0112156;%%

\bibitem{eeKout} 
A.~Banfi, G.~Marchesini, Yu.~L.~Dokshitzer and G.~Zanderighi,
%``QCD analysis of near-to-planar 3-jet events,''
JHEP {\bf 0007}, 002 (2000);
%[arXiv:hep-ph/0004027].
%%CITATION = HEP-PH 0004027;%%
%A.~Banfi, Yu.~L.~Dokshitzer, G.~Marchesini and G.~Zanderighi,
%``QCD analysis of D-parameter in near-to-planar three-jet events,''
JHEP {\bf 0105}, 040 (2001).
%[arXiv:hep-ph/0104162].
%%CITATION = HEP-PH 0104162;%%

\bibitem{MP} 
D.~H.~Bailey, ``A Portable High Performance
Multiprecision Package'', NASA Ames RNR Technical Report RNR-90-022; 
``A Fortran-90 Based Multiprecision System'', RNR Technical Report
RNR-94-013. 

\bibitem{D0}
I.~A.~Bertram  [D0 Collaboration],
%``Jet Results At The D0 Experiment,''
Acta Phys.\ Polon.\ B {\bf 33}, 3141 (2002).
%%CITATION = APPOA,B33,3141;%%

\bibitem{NLOJET}
Z.~Nagy,
%``Three-jet cross sections in hadron hadron collisions at next-to-leading  order,''
Phys.\ Rev.\ Lett.\  {\bf 88}, 122003 (2002);
%[arXiv:hep-ph/0110315].
%%CITATION = HEP-PH 0110315;%%
%Z.~Nagy,
%``Next-to-leading order calculation of three-jet observables in hadron hadron collision,''
hep-ph/0307268.
%%CITATION = HEP-PH 0307268;%%

%\bibitem{cFAnl} It can be automatically established that $\tau_\perp$
%  is additive, $V(\{\tilde p\}, k_1\ldots k_m) = \sum_i V(\{\tilde
%  p\}, k_i)$, meaning that one can use the known result $\cF =
%  e^{-\gamma_E R'}/\Gamma(1+R')$ \cite{CTTW}.

\bibitem{CTEQ}
J.~Pumplin \textit{et al.},
% D.~R.~Stump, J.~Huston, H.~L.~Lai, P.~Nadolsky and W.~K.~Tung,
%``New generation of parton distributions with uncertainties 
%from global  QCD analysis,''
JHEP {\bf 0207}, 012 (2002). 
%[arXiv:hep-ph/0201195].
%%CITATION = HEP-PH 0201195;%%


\end{thebibliography}
\end{document}